# Infrared Surface Brightness Distances to Cepheids: a comparison of Bayesian and linear-bisector calculations

Thomas G. Barnes III[1], Jesper Storm[2], William H. Jefferys[3], Wolfgang P. Gieren[4] and Pascal Fouqué[5]

[1] The University of Texas at Austin, McDonald Observatory, 1 University Station, C1402, Austin, TX, USA 78712–0259
[2] Astrophysikalisches Institut Potsdam, An der Sternwarte 16, D-14482 Potsdam, Germany
[3] The University of Texas at Austin, Dept. of Astronomy, 1 University Station, C1400, Austin, TX, USA 78712–0259
[4] Universidad de Concepción, Departamento de Física, Casilla 160-C, Concepción, Chile
[5] Observatoire Midi-Pyrénées, Laboratoire d'Astrophysique (UMR 5572), 14, avenue Edouard Belin, F-31400 Toulouse, France e-mail: `tgb@astro.as.utexas.edu`

**Abstract.** Bayesian statistical calculations and linear-bisector calculations for obtaining Cepheid distances and radii by the infrared surface brightness method have been compared for a set of 38 Cepheids. The distances obtained by the two techniques agree to $1.5\% \pm 0.6\%$ and the radii agree to $1.1\% \pm 0.7\%$. Thus the two methods yield the same distances and radii at the $2\sigma$ level. This implies that the short distance to the LMC found in recent linear-bisector studies of Cepheids is not a result of simplifications in the mathematical approach. The computed *uncertainties* in distance and radius are larger in the Bayesian calculation typically by a factor of three.

## 1. Motivation

The infrared surface brightness technique (ISB method) is a powerful method for determining distances to Cepheid variables. It is independent of other astrophysical distance scales, nearly independent of errors in reddening, and may be applied to arbitrarily-chosen, individual Cepheids at great distances. Importantly the surface brightness relation has recently been put on a solid foundation by means of interferometric angular diameters of Cepheids (Kervella et al. 2004, Nordgren et al. 2002).

However, solutions to the surface brightness equations have been criticized as not mathematically rigorous (Laney & Stobie 1995, Barnes & Jefferys 1999). The specific issues that arise concern handling the errors-in-variables problem, non-objective model selection, and deficiencies in the error propagation through the radial velocity integral. These issues lead to the possibility that the distances, radii, and their uncertainties may be systematically in error.

*Send offprint requests to*: T. G. Barnes

Investigating the possibility of systematic errors in ISB distances is timely as Storm et al. (2005) have found a short distance to the LMC based on a linear-bisector analysis of six Cepheids in the LMC cluster NGC1866. They also found a slope to the LMC Cepheid period-luminosity relation that is substantially different than found in the OGLE-II magnitudes (Udalski et al. 1999). It is important to determine if their results are affected by the mathematical method used in their analysis.

Barnes et al. (2003) developed a Bayesian Markov Chain Monte Carlo (MCMC) solution to the surface brightness equations that is mathematically rigorous. Here we do a direct comparison between the Bayesian MCMC solutions and the linear-bisector calculations for 38 Cepheids to discover any differences. The two calculations are based on identical data, identical surface brightness equations and identical physical constants to ensure that only the mathematical approaches are compared.

## 2. Data

Storm et al. (2004) studied 34 Galactic Cepheids for distances and radii using the $V_0, (V-K)_0$ ISB method. Gieren et al. (2005) enlarged the sample with four more stars. We have used the combined set of 38 Cepheids. The stellar data provided by these studies are the photometric measures $V, (V-K)$, radial velocities $V_r$, pulsation period $P$, and pulsation phases $\theta$.

We adopted $E(B-V)$ from Fernie's (1990) tabulation for Cepheids. We used a different reddening law than used in the linear-bisector studies: $A_V = 3.26E(B-V)$ and $E(V-K) = 2.88E(B-V)$. The linear-bisector calculations were repeated with the new reddening law for all 38 Cepheids.

## 3. The Linear-bisector Computation

The ISB relation adopted in the linear-bisector analysis and here is the one determined by Fouqué & Gieren (1997):

$$F_V = 3.947 - 0.131(V-K)_0 \quad (1)$$

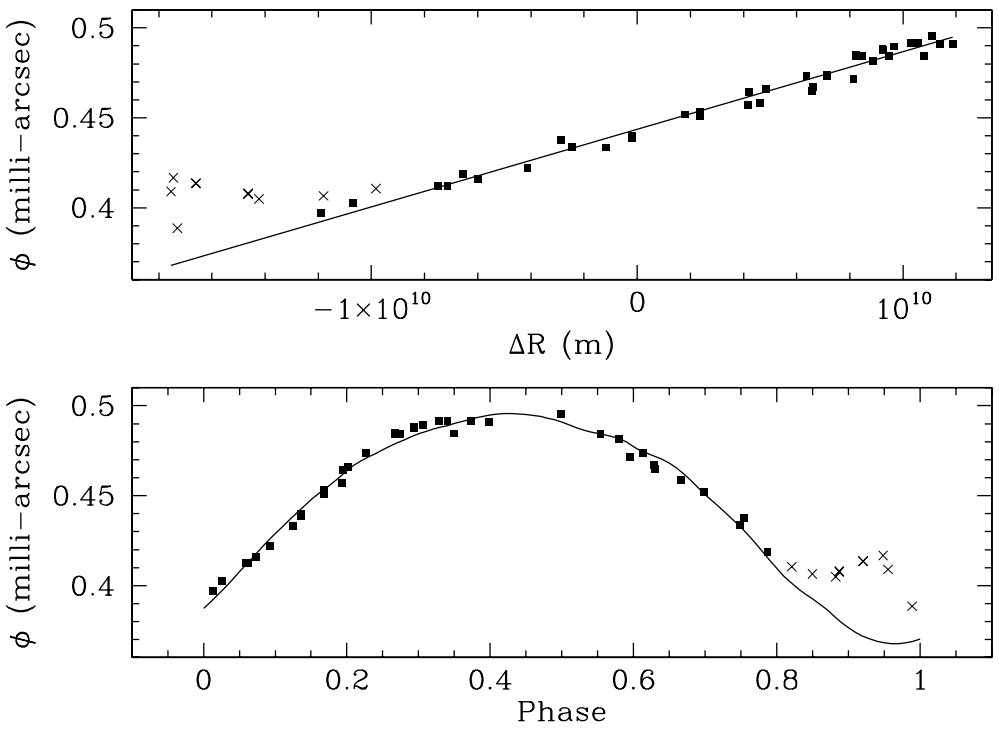

**Fig. 1.** Points represent the photometrically determined angular diameters for AQ Pup. Crosses represent values that were eliminated from the fit. The line in the upper panel shows the bisector fit. The curve in the lower panel shows the angular diameters obtained from integrating the radial velocity curve at the derived distance.

In the linear-bisector analysis equation (2) is solved for the angular diameter $\phi(t)$:

$$4.2207 - 0.1V_0(t) - 0.5\log\phi(t) = A + B(V-K)_0(t) \quad (2)$$

where $A, B$ are taken from equation (1).

The radial velocity data $V_r(t)$ are linearly interpolated in phase and equation (3) integrated to obtain the linear radius variation, $\Delta R(t)$:

$$\Delta R(t) = -\int p\left(V_r(t) - V_\gamma\right) dt \quad (3)$$

We adopted a relation between $p$ and period developed by Gieren et al. (1989). Note that the choice of $p$ is irrelevant as it will cancel out in our comparison between the methods.

Equation (4) was then solved for the distance $r$ and the mean angular diameter $\phi_0$:

$$\Delta R(t) = r(\phi_0 + \phi(t))/2000 \quad (4)$$

As illustrated in Figure 1, for many of the stars the angular diameter variation was a very poor match to the displacement curve in the phase interval 0.8 − 1.0, Therefore this phase interval was ignored for all stars.

Some Cepheids showed a small phase shift between the photometry and the radial velocities. This shift was measured and removed.

## 4. The Bayesian Markov Chain Monte Carlo Computation

The Bayesian MCMC method that we applied is described in detail by Barnes et al. (2003). The model for the infrared surface brightness calculation is developed by rearranging equation (2) as follows:

$$(V-K)_0(t) = \frac{1}{B}(4.2207 - 0.1V_0(t) - A) - \frac{1}{B}(0.5\log(\phi_0 + 2000\Delta R(t)\pi)) \quad (5)$$

where we have replaced $1/r$ with $\pi$, the parallax in arcseconds. Within this model the likelihood function is specified in a straightforward way (in eq. [12] of Barnes et al. 2003). Priors are adopted for all parameters in the problem. We modeled the photometry and the radial velocity data as drawn from normal distributions with variances given by the observational uncertainties. The time variations of the photometry and radial velocities are modeled by Fourier series on the pulsation phase with orders to be determined. We ignored the pulsation phase interval $0.8 - 1.0$. We also fixed the phase shift at the value determined in the linear-bisector calculation. We ran 10,000 samples in the MCMC calculation after a burn-in of 1,000 samples.

To ensure that we used identical data in both calculations, the Bayesian analysis used the same input data files as used in the linear-bisector analysis.

Posterior probability distributions were determined for all the model parameters of interest. We converted the parallax and mean angular diameter results into distance and mean linear diameter. The distance result for U Sgr is shown in Fig. 2.

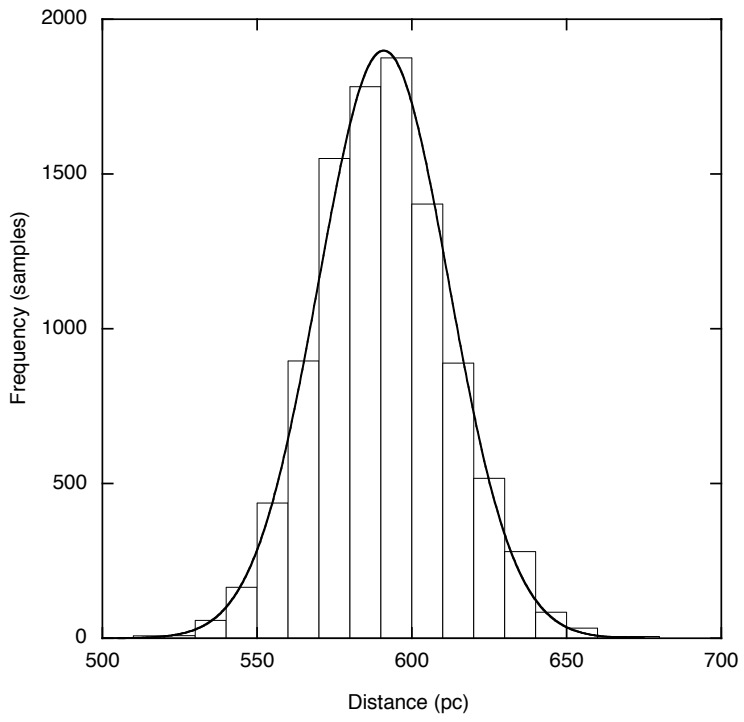

**Fig. 2.** The posterior probability distribution for the distance to U Sgr is shown as a histogram. A normal distribution of the same mean distance, sigma and area is over-plotted.

## 5. Comparison of the Results

Figure 3 shows the Bayesian distances plotted against the linear-bisector distances. (The radius plot is similar.) There is no obvious difference between the values from the two calculations. To look for subtle differences, we computed the ratio of the Bayesian distance to the linear-bisector distance and looked for correlations with *log*(*P*), distance, radius, ... No correlations were found. In a weighted comparison, the distance calculations differ by $1.5\% \pm 0.6\%$, and the radius calculations differ by $1.1\% \pm 0.7\%$, both with the Bayesian values being larger.

On the other hand, the *uncertainties* in the distances and radii disagree substantially between the two calculations. As shown in Fig. 4 the Bayesian uncertainty is typically three times the linear-bisector uncertainty, although with much variation from star to star. Because the linear-bisector calculation is known to have mathematical deficiencies that can affect the uncertainties, and the Bayesian MCMC does not have these deficiencies, we attribute the substantial differences in the uncertainties to underestimation of those quantities in the linear-bisector method.

## 6. Conclusions

We set out to determine whether ISB estimates of Cepheid distances and radii by the linear-bisector calculation are affected by the known mathematical shortcomings of that calculation. Based on comparison of Bayesian MCMC and linear-bisector calculations, we find that the distances and radii are *not* adversely affected at the $2\sigma$ level but that the uncertainties in these quantities are seriously underestimated in the linear-bisector calculation by a typical factor of three. That the uncertainties of the linear-bisector method are underestimated has also been found by Storm et al. (2005). In their work on six Cepheids in the LMC cluster NGC1866, they find the scatter in the distances to be twice the formal uncertainty, in line with the results here.

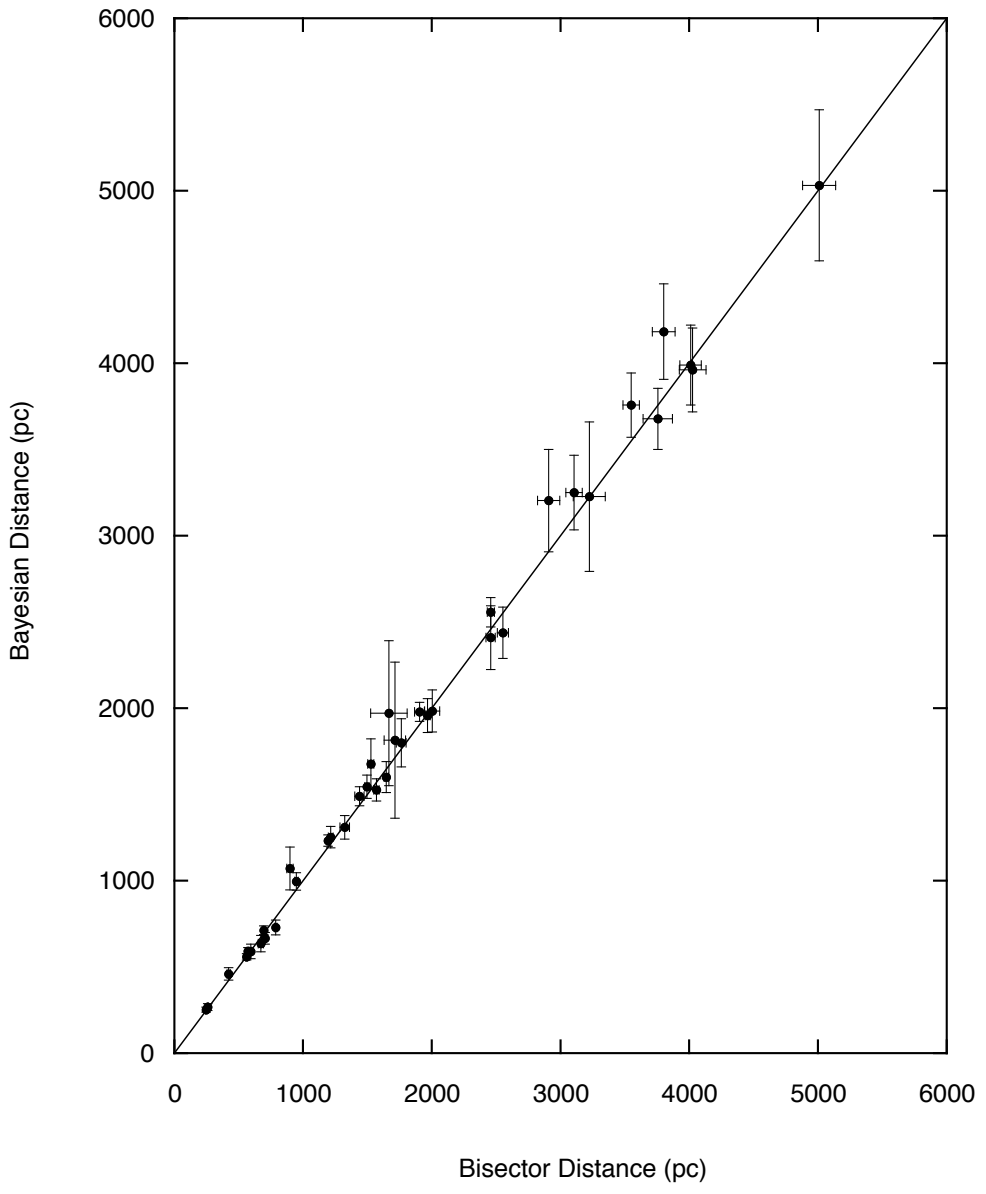

**Fig. 3.** The Bayesian distances are plotted against the linear-bisector distances, each with its $1\sigma$ error bar. For nearby Cepheids the uncertainties are smaller than the symbols. The identity line is shown for reference.

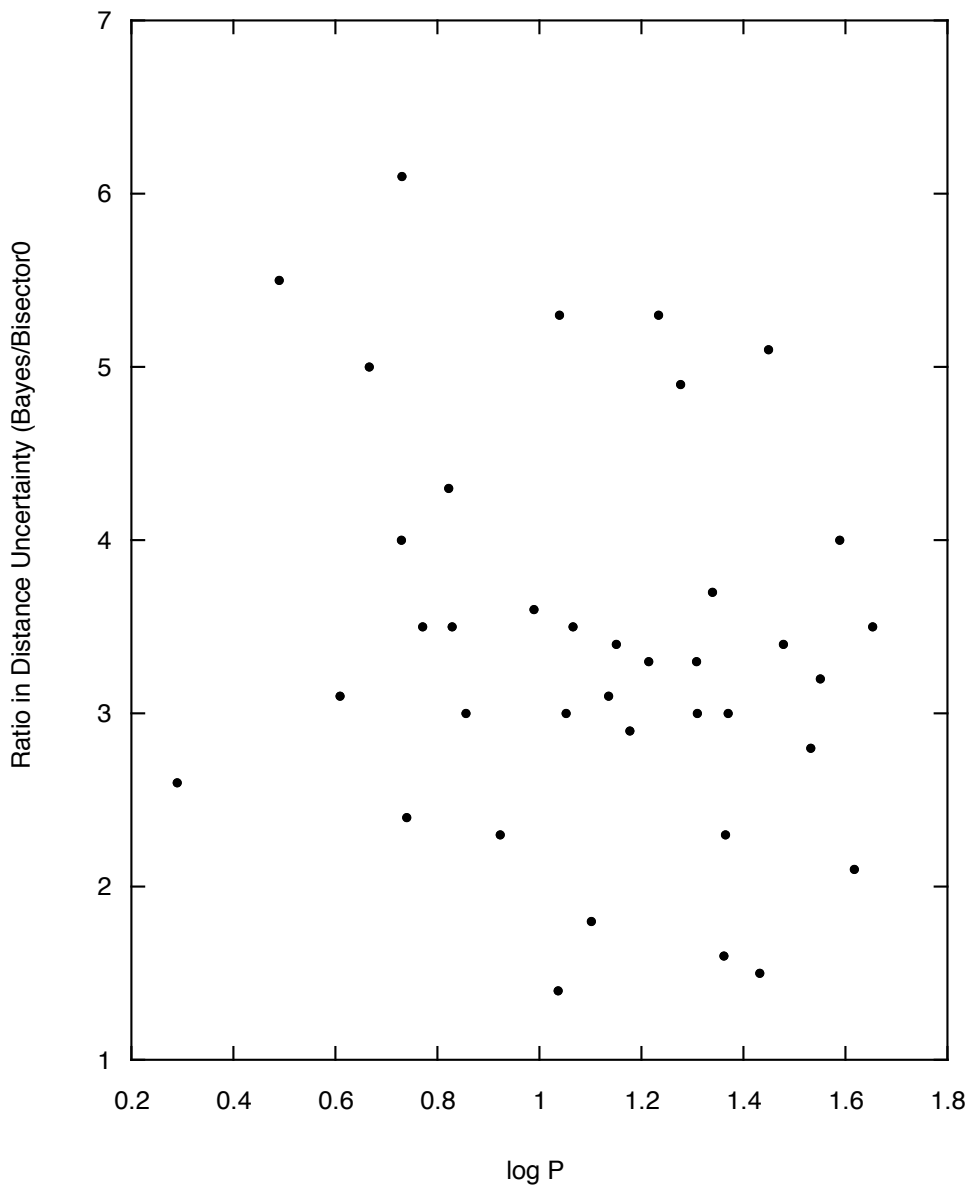

**Fig. 4.** The ratio of the Bayesian distance uncertainty to the bisector distance uncertainty for the same Cepheid is plotted against *logP*.

*Acknowledgements.* WPG acknowledges financial support for this work from the Chilean Center for Astrophysics FONDAP 15010003.

## References

Barnes, T. G., & Jefferys, W. J. 1999, in ASP Conf. Ser. 167, Harmonizing Cosmic Distance Scales in a Post-*Hipparcos* Era, ed. D. Egret & A. Heck, (San Francisco:ASP), 243

Barnes, T. G., Jefferys, W. H., Berger, J. O., Mueller, P. J., Orr, K., & Rodriguez, R. 2003, ApJ, 592, 539; erratum ApJ, 611, 621, 2004

Fernie, J. D. 1990, ApJS, 72, 153

Fouqué, P., & Gieren, W. P. 1997, A&A, 320, 799

Gieren, W. P., Barnes, T. G., & Moffett, T. M. 1989, ApJ, 342, 467

Gieren, W. P., Storm, J., Barnes, T. G., Fouqué, P. Pietrzyński, G., & Kienzle, F. 2005, ApJ, 627, 224

Kervella, P., Bersier, D., Mourard, D., Nardetto, N., Fouqué, P., & Coudé du Foresto, V. A&A, 428, 587

Laney, C. D., & Stobie, R. S. 1995, MNRAS, 274, 337

Nordgren, T. E., Lane, B. F., Hindsley, R. B., & Kervella, P. 2002, AJ, 123, 3380

Storm, J., Carney, B. W. , Gieren, W. P. , Fouqué, P. , Latham, D. W., & Fry, A.M. 2004, A&A, 415, 531

Storm, J., Gieren, W. P. , Fouqué, P. , Barnes, T. G., & Gómez, M. 2005, A&A, in press

Udalski, A., Soszyński, I., Szymański, M., Kubiak, M., Pietrzyński, G., Woźniak, P., & Żebruń, K. 1999, Acta Astron., 49, 223